\documentclass[prl,twocolumn,showpacs,preprintnumbers,amsmath,amssymb]{revtex4}

\usepackage{graphicx}
\usepackage{dcolumn}
\usepackage{bm}

\begin{document}

\title{Mesoscopic field and current compensator based on a hybrid superconductor-ferromagnet structure}

\author{M. V. Milo\v{s}evi\'{c}}
\author{G. R. Berdiyorov}
\author{F. M. Peeters}
\email{Francois.Peeters@ua.ac.be}

\affiliation{Departement Fysica, Universiteit Antwerpen,
Groenenborgerlaan 171, B-2020 Antwerpen, Belgium}

\date{\today}

\begin{abstract}
A rather general {\it enhancement} of superconductivity is
demonstrated in a hybrid structure consisting of submicron
superconducting (SC) sample combined with an in-plane ferromagnet
(FM). The superconducting state resists much \textit{higher}
applied magnetic fields for \textit{both perpendicular
polarities}, as applied field is screened by the FM. In addition,
FM induces (in the perpendicular direction to its moment) two
opposite current-flows in the SC plane, under and aside the
magnet, respectively. Due to the compensation effects,
superconductivity persists up to {\it higher applied currents}.
With increasing current, the sample undergoes
SC-``resistive''-normal state transitions through a mixture of
vortex-antivortex and phase-slip phenomena.
\end{abstract}


\maketitle

Over the last years, the superconductor-ferromagnet (SC-FM)
hybrids received a lot of attention as one of the rare systems
where ferromagnetism and singlet superconductivity coexist (for
review, see \cite{pokr}). These hybrid structures are looked upon
as candidates for futuristic nanoelectronics, combining
superconducting circuits with magnetic storage elements. As better
understanding is needed, the ongoing studies are mainly focused on
fundamental properties of nanoscale SC-FM samples and plethora of
related phenomena.

For example, although ferromagnetism in general suppresses
superconductivity, direct SC-FM coupling appears to be crucial for
the $\pi$-phase state with the critical current inversion in
SC-FM-SC junctions \cite{ryaz} and Josephson current enhancement
in SC-FM tunnel structures with very thin FM layers \cite{berg}.
On the other hand, the nontrivial interplay between magnetism and
superconductivity can be achieved even if SC and FM are not
electronically coupled, as they still interact through the
emerging magnetic fields. In that respect, arrays of submicron
magnetic particles are used for applying well-defined local
magnetic fields in the underlying superconductor \cite{martin}.
One of the first applications of these nano-magnets was to
engineer the pinning force of superconducting films, such that the
critical current $j_{c}$ as a function of applied magnetic field
is increased due to a collective locking of the flux lattice to
the magnetic array \cite{martin,morgan}. Since then, because of
the technological relevance, enhancement of critical parameters in
SC-FM heterostructures is of vast theoretical and experimental
interest. Genenko {\it et al.} predicted theoretically an
increased edge barrier critical current in superconductors
completely surfaced by magnetic material \cite{genenko}. In that
case, a demagnetized magnetic layer acts as a magnetic screen,
effectively shielding the Meissner state. Two years ago, Lange
{\it et al.} measured higher critical field in SC films regularly
structured by out-of-plane magnetized dots \cite{lange}. However,
this behavior strongly depended on the polarity of the applied
field $\textbf{H}_{ext}$: for given FM-magnetization $\textbf{M}$,
an enhancement of the critical parallel field
($\textbf{H}_{ext}\parallel \textbf{M}$) was achieved at expense
of the antiparallel one. The same behavior was found both
experimentally \cite{golub} and theoretically \cite{miskoold} in
mesoscopic SC disks with out-of-plane FM dot on top.

\begin{figure}[b]
\includegraphics[height=4.2cm]{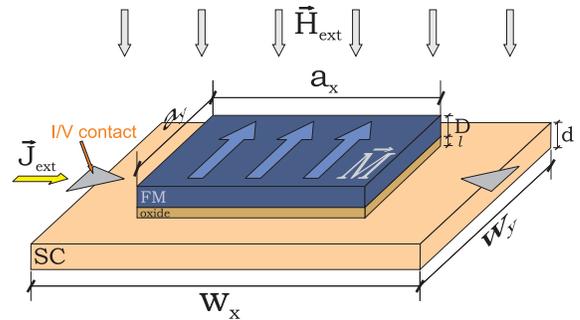}
\caption{\label{fig:fig1}A superconducting (SC) sample underneath
a square ferromagnetic (FM) dot with in-plane magnetization
$\textbf{M}$ (separated by an insulating layer). Depicted
directions of $\textbf{M}$, the applied external magnetic field
$\textbf{H}_{ext}$ and current $\textbf{j}_{ext}$ are denoted as
\textit{positive} throughout the article.}
\end{figure}

The first objective of the present Letter is to design a SC-FM
hybrid structure where most critical properties can be tailored
practically at will. For that matter, we consider a thin submicron
superconducting sample with a ferromagnetic dot with in-plane
magnetization on top (see Fig. \ref{fig:fig1}). Such a device
realization offers full exploitation of the magnetic flux pinning
\cite{margriet}, dynamical properties of mesoscopic
superconductors \cite{ivlev}, and related vortex-antivortex
phenomena \cite{misko}. Due to the opposite magnetic field at the
poles of the magnet, the field-compensation effects lead to the
critical field enhancement for both positive and negative applied
field (see Fig. \ref{fig:fig1}). At the same time, our SC-FM
sample might act as a current compensator as well: as a novel
concept, the applied current is met by opposing FM-induced
supercurrents, resulting in a larger critical current.

In our theoretical treatment of this system, we rely upon the
Ginzburg-Landau (GL) formalism. In, the stationary case, we solve
self-consistently two GL equations, derived from the Gibbs energy
functional. For all details of this approach, we refer to Refs.
\cite{misko,schweigert1}.

To understand the dynamical properties of the device, we studied
the current-voltage characteristics using the time-dependent
Ginzburg-Landau (TDGL) equation \cite{Kramer}
\begin{eqnarray}
\frac{u}{\sqrt{1+\Gamma^2|\Psi|^2}} \left(\frac {\partial
}{\partial t} +{\rm i}\varphi +
\frac{\Gamma^2}{2}\frac{\partial|\Psi|^2}{\partial t}
\right)\Psi= \nonumber \\
=(\nabla - {\rm i} {\bf A})^2 \Psi +(1-T-|\Psi|^2)\Psi,
\label{tdgl1}
\end{eqnarray}
coupled with the equation for the electrostatic potential $\Delta
\varphi = {\rm div}\left({\rm Im}(\Psi^*(\nabla-{\rm i}{\bf
A})\Psi)\right)$. Here, the distance is measured in units of the
coherence length $\xi(0)$, $\Psi$ is scaled by its value in the
absence of magnetic field $\psi_{0}$, time by $\tau_{GL}(0)=\pi
\hbar \big/8k_BT_cu$, vector potential ${\bf A}$ by $c\hbar\big/
2e\xi(0)$, and the electrostatic potential by
$\hbar\big/2e\tau_{GL}(0)$. $\Gamma=2\tau_E \psi_{0}/\hbar$, with
$\tau_E$ being the inelastic electron-collision time. For Al
samples $\tau_E\sim 10$ns, which results in $\Gamma\approx 1000$.
Parameter $u=5.79$ is taken from Ref. \cite{Kramer}. Note that in
Eq. (\ref{tdgl1}) the screening of the magnetic field is
neglected, as we restrict ourselves to thin SC samples ($d<\xi$).
The points where external current $j_{ext}$ is injected in the
sample (see Fig. \ref{fig:fig1}) were simulated as normal
metal-superconductor contacts, i.e. with $\Psi=0$ and $-\nabla
\varphi=j_{ext}$. At the remainder of the sample edges, Neumann
boundary condition was used ($j_{\bot}=0$).

\begin{figure}[t]
\includegraphics[height=6.3cm]{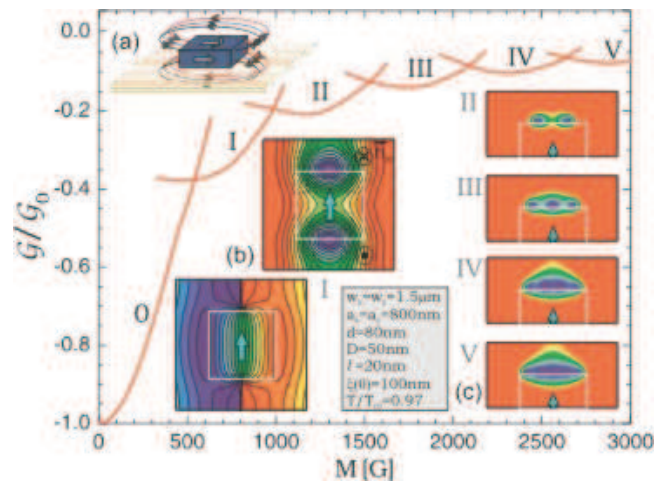}
\caption{\label{fig:fig2}The Gibbs free energy diagram
($\mathcal{G}_{0}=H_{c}^{2}V/8\pi$) as a function of
FM-magnetization. Roman numbers denote number of FM-induced
vortex-antivortex pairs. Inset (a) illustrates the FM-stray field
lines; (b) the Cooper-pair density [upper figure, darkest color -
zero density (online blue/red-low/high density)] and
superconducting phase contourplot [gradation of grey color shows
the circulation of phase $0-2\pi$ (online blue/red-0/2$\pi$
phase)] for state I; (c) the $|\psi|^{2}$-density plots (scale
adjusted for clarity) under the positive pole of the magnet, for
states II to V (FM-edge depicted by white lines).}
\end{figure}

We consider a square $Al$ sample with parameters easily achievable
with modern lithographic techniques: size $a_{x}=a_{y}=1.5\mu$m,
thickness $d=80$nm, separated by an oxide layer of thickness
$l=20$nm from the square FM with size $w_{x}=w_{y}=800$nm and
thickness $D=50$nm. The SC material is characterized by its
coherence length at zero temperature, which we take as
$\xi(0)=100$nm (typical value for mesoscopic $Al$ samples
\cite{golub}), and FM material by its saturation magnetization
$M$. In our calculations, the FM is positioned at the center of
the SC square, and since its stray field has opposite polarity at
the FM-poles (see inset (a) in Fig. \ref{fig:fig2}), the total
flux $\Phi_{FM}$ penetrating the SC equals zero. This feature
inevitably leads to the appearance of vortex-antivortex
configurations for sufficiently strong magnetization $M$
\cite{misko}.

Fig. \ref{fig:fig2} shows the Gibbs free energy \cite{schweigert1}
of the superconducting state, obtained after sweeping up/down the
FM-magnetization, where the number of induced vortex-antivortex
(VAV) pairs is denoted by Roman numbers. Note that these VAV
states are the first found vortex states with zero total vorticity
in finite mesoscopic SC samples. Insets (b,c) in Fig.
\ref{fig:fig2} show the $|\psi|^{2}$-density plots of successive
VAV states. As vortices and antivortices are confined at the
FM-poles (where the stray field is maximal), they are effectively
kept apart by the FM. In other words, the superconducting region
under the FM always remains (anti)vortex-free. As a result,
superconductivity can be sustained in the sample up to very large
FM-magnetization (as the slope $\partial \mathcal{G}/\partial M$
decreases in Fig. \ref{fig:fig2}). Note that this is not the case
if FM has perpendicular magnetization, when total flux $\Phi_{FM}$
captured by SC is positive and FM-induced vortices destroy
superconductivity in the heart of the sample
\cite{golub,miskoold}. The experiment of Ref. \cite{golub}
revealed that when such a sample is exposed to homogeneous
external field $H_{ext}$, the $H_{ext}-T$ boundary is shifted
towards positive fields due to the compensation with $\Phi_{FM}$,
resulting in higher positive critical field (and consequently
reduced negative one).

\begin{figure}[t]
\includegraphics[height=6.3cm]{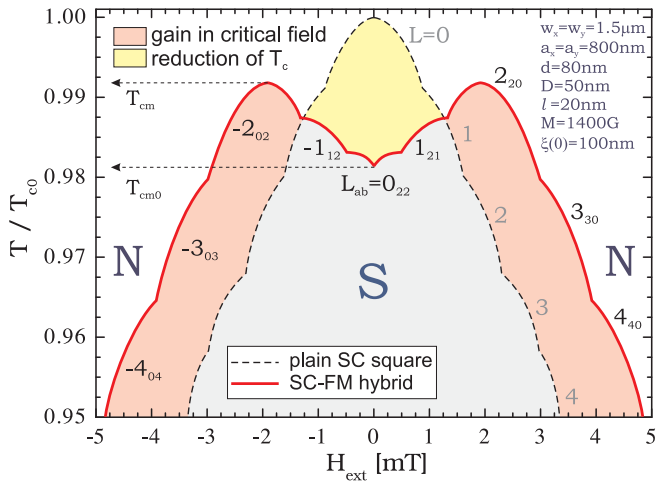}
\caption{\label{fig:fig3}The superconducting/normal (S/N) phase
boundary for a square SC sample, without (dashed curve) and with
(solid curve) an in-plane FM dot on top. The total vorticity of
the sample is denoted by $L$, while indices $a,b$ show the number
of vortices and antivortices (induced and/or pinned by FM) at
corresponding FM poles, respectively ($L=a-b$).}
\end{figure}

In Fig. \ref{fig:fig3}, the $H_{ext}-T$ superconducting/normal
(S/N) phase boundary of our sample is shown, in the case of FM
with bulk Co magnetization $M=1400$G (solid line), compared to the
case without FM (dashed line). The S/N boundary exhibits {\it
three novel features}: (i) {\bf the $\mathcal{M}$-shaped boundary}
- the critical temperature for $H_{ext}=0$ is reduced ($T_{cm0}$),
and is maximal for two symmetric non-zero $H_{ext}$ values
($T_{cm}$); (ii) for $T_{cm0}<T<T_{cm}$ the {\bf N-S-N-S-N
multi-reentrant behavior} is observed during $H_{ext}$ sweep; and
(iii) {\bf substantial critical field enhancement} is found for
{\it both} $H_{ext}$ polarities.

The physical reason for these phenomena lies in the magnetic field
compensation. In this particular case, for $H_{ext}=0$, the S/N
transition at $T=T_{cm0}$ occurs for 2 vortex-antivortex pairs
induced by the FM (state II in Fig. \ref{fig:fig2}). Although
their centers are pin-pointed at the FM-poles, these
(anti)vortices are covering the whole sample as $\xi(T)$ becomes
large. When $H_{ext}>0$ is applied, the external flux is
``absorbed'' by the FM-induced antivortex. This effectively
recovers superconductivity, and increases $T_{c}$. Each kink in
the Little-Parks-like S/N boundary with increasing $H_{ext}$
corresponds to a change in total vorticity of $\Delta L=1$, where
external flux lines are first annihilated by FM-induced
antivortices, and in the absence of antivortices pinned on the
positive pole of the FM, where the stray field and $H_{ext}$ are
aligned (as found in Refs. \cite{margriet}). In the latter case,
each additional vortex suppresses superconductivity and $T_{c}$
decreases. However, the SC state remains protected at the opposite
pole of the magnet which results in significantly higher critical
field. In Fig. \ref{fig:fig3}, an enhancement as high as $\sim
40\%$ is achieved. This percentage can be even larger, if stronger
magnetic materials are used (note high $M_{cr}$ in Fig.
\ref{fig:fig2}). The reduced zero-field critical temperature
$T_{cm0}$ is only within few percent from $T_{c0}$, but maximal
$T_{cm}$ (at $H_{ext}=1.97$mT) is several percent higher than the
corresponding $T_{c}$ value in the absence of the FM.

\begin{figure}[b]
\includegraphics[height=9.5cm]{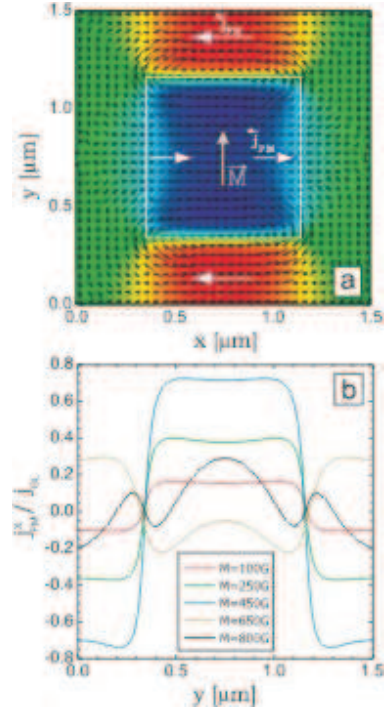}
\caption{\label{fig:fig4}(a) Vectorplot of the SC-current $j_{FM}$
induced by an in-plane FM (Meissner state), superimposed on the
contourplot of $j_{FM}^{x}$ component. (b) The
$j_{FM}^{x}$-profile in the central cross-section for different
FM-magnetization ($T=0.97T_{c0}$).}
\end{figure}
For $H_{ext}<0$, the scenario is completely analogous, and S/N
boundary is therefore symmetric. Note that this symmetry directly
leads to feature (ii), which is actually a very rare magnetic
field-\textit{induced}-superconductivity (FIS) phenomenon. Similar
unconventional behavior was reported earlier for materials like
(EuSn)Mo$_6$S$_8$ and $\lambda$-(BETS)$_{2}$FeCl$_{4}$, and for SC
films with out-of-plane FM-arrays on top \cite{lange}. However,
our sample holds a unique property - FIS is achieved for {\it
both} perpendicular polarizations of applied field $H_{ext}$ (e.g.
Fig. \ref{fig:fig3} for $T=0.985T_{c0}$).

\begin{figure*}[t]
\includegraphics[height=4.8cm]{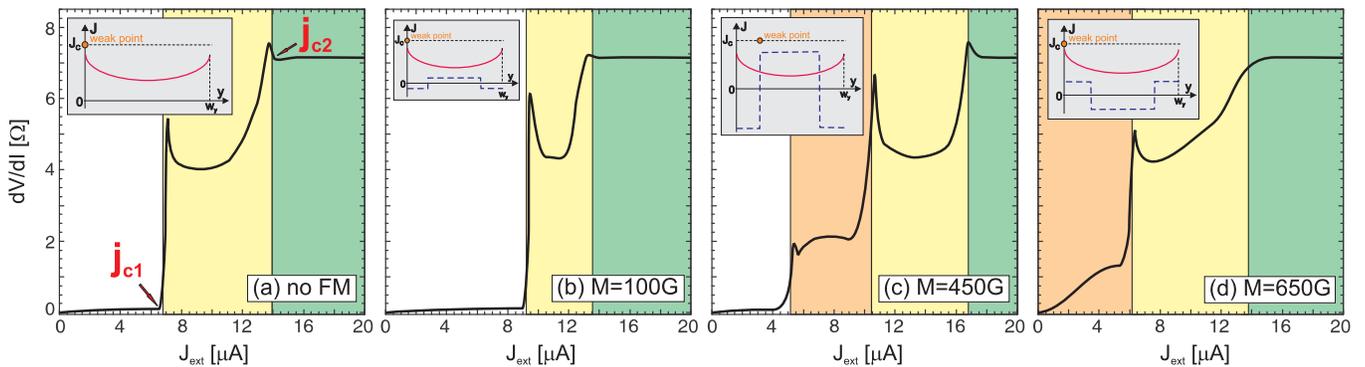}
\caption{\label{fig:fig5}Calculated differential resistance of the
sample as function of the applied DC current, for different
FM-magnetization ($T=0.97T_{c0}$). White areas denote Meissner
state, light grey (yellow online) - ``resistive'', medium grey
(orange) - vortex-antivortex, and dark grey (green) - normal
state. Insets illustrate schematically the distribution of the
applied (solid line) and FM-induced current ($j_{FM}^{x}$
component, dashed line) across the sample.}
\end{figure*}
Obviously, the above described phenomena are directly related to
the strongly inhomogeneous FM-stray field (with zero average). Yet
another interesting feature can be found in the
stray-field-induced currents ${\bf j}_{FM}$. Due to the field
landscape, these currents are actually circulating around the
poles of the magnet, which ultimately results in two opposite
current flows, under and aside the FM, in a direction
perpendicular to the FM-polarization (see Fig. \ref{fig:fig4}(a)).
Obviously, with increasing FM-magnetization, the amplitudes of
${\bf j}_{FM}$ grow (as shown in Fig. \ref{fig:fig4}(b)), until
the depairing current is reached and vortex-antivortex (VAV) pair
nucleates at the FM-poles. Consequently, $j_{FM}$ completely
reverses (Fig. \ref{fig:fig4}(b), yellow line), but changes
polarity again with increasing $M$, before the appearance of the
following VAV pair (see Fig. \ref{fig:fig2}). This dual, step-like
$j_{FM}$ profile may strongly affect the response of the device on
the applied current in the $x$-direction (see Fig.
\ref{fig:fig1}). In order to investigate the critical current and
dynamical properties of the system, we employ TDGL formalism. The
key results are shown in Fig. \ref{fig:fig5}, as differential
resistance (obtained from calculated I-V characteristics) as
function of applied current $j_{ext}$. Two critical currents,
denoted as $j_{c1}$ and $j_{c2}$ in Fig. \ref{fig:fig5}, can be
identified. $j_{c1}$ is the current at which the sample loses its
zero-resistance and transits to the so-called ``resistive'' state.
$j_{c2}$ has the more conventional meaning of the current at which
the SC state becomes unstable.

When external current is applied to a plain SC square (Fig.
\ref{fig:fig5}(a)), it is non-uniformly distributed in the sample,
with its maxima at the side-edges (see cartoon in the inset). It
is at these weak points where the vortex nucleates when the
depairing current is reached (for corresponding $j_{ext}=j_{c1}$).
Due to the Lorentz force, this vortex is expelled across the
sample, and subsequently nucleating again. To prevent the
destruction of superconductivity, the SC phase exhibits a jump of
$2\pi$ at this ``phase-slip line'' \cite{denis}. This phase-slip
can be theoretically interpreted as an infinitely-fast moving
vortex. With continuous increase of $j_{ext}$, the current further
suppresses the order parameter at the contacts (see Fig.
\ref{fig:fig1}) finally establishing a normal path between them
(for $j_{ext}=j_{c2}$) and normal-metal resistance is reached in
Fig. \ref{fig:fig5}(a). Note that the presence of S/N contacts in
our simulation gives small, but finite resistance even for
$j_{ext}<j_{c1}$.

In the case of our SC-FM device, for small $M$, the FM induces
current opposite to the external one, along the sample edges.
Therefore, the resulting current at the weak points is decreased,
and the critical current for vortex entry (phase-slip and non-zero
resistance) {\it increases}. Fig. \ref{fig:fig5}(b) shows
exquisite $j_{c1}$ enhancement of $\sim 35\%$, for $M=100$G (see
Fig. \ref{fig:fig4}(b) for $j_{FM}$ profile, and Ref.
\cite{denis1}). However, for larger FM-magnetization, after
superposition of $j_{ext}$ and $j_{FM}$, although edge current is
further suppressed, the depairing current may be reached under the
FM where $j_{FM}$ is maximal. As a result, VAV pair nucleates, and
SC transits to the resistive state for a low $j_{c1}$ value (Fig.
\ref{fig:fig5}(c)). Following VAV creation, $j_{FM}$ changes
polarity, and now compensates $j_{ext}$ between the contacts
instead of at the edges. Eventually, with further increase of
$j_{ext}$, $j_{FM}$ is overwhelmed by the applied current, and VAV
pair is expelled from the sample; the influence of FM becomes
negligible as further scenario resembles the one of Fig.
\ref{fig:fig5}(a): current is again maximal at the edge,
phase-slip occurs, and superconductivity is destroyed. However,
due to described VAV nucleation and current compensation between
the contacts, for $M=450$G we obtained a {\it remarkable
enhancement} of $j_{c2}$ of $\sim 21.5\%$ (Fig.
\ref{fig:fig5}(c)).

For higher $M$, FM may induce a VAV pair in the sample (Fig.
\ref{fig:fig2}(b)). In that case, even for very low applied
current, a finite resistance was found ($j_{c1}=0$). This feature
can serve as a tool for experimental detection of VAV pairs in
contrast to the Meissner state in SC-FM hybrids. For certain value
of $j_{ext}$, vortex and antivortex are depinned and leave the
sample, followed by an immediate phase-slip and consequent
transition to the normal state. Due to the absence of the
zero-resistance state, both $j_{c1}$ and $j_{c2}$ are
significantly decreased (Fig. \ref{fig:fig5}(d)).

In conclusion, we proposed a SC-FM device where {\it both}
critical field and critical current can be substantially enhanced.
Although our dynamical simulations are valid only in close
vicinity of $T_{c}$, the main idea is generally applicable.
Detailed influence of parameters and different dynamic regimes
will be analyzed in a separate article.

This work was supported by the Flemish Science Foundation
(FWO-Vl), the Belgian Science Policy, and the JSPS/ESF-NES
program. Discussions with D.Y. Vodolazov are gratefully
acknowledged.

\end{document}